# Cyber Pirates Ahoy!
# An Analysis of Cybersecurity Challenges in the Shipping Industry

George Grispos and William R. Mahoney

*School of Interdisciplinary Informatics*
*University of Nebraska-Omaha*
*Omaha, NE, United States*

*E-mail:* ggrispos@unomaha.edu;
wmahoney@unomaha.edu

**Abstract**: *Maritime shipping has become a trillion-dollar industry that now impacts the economy of virtually every country around the world. It is therefore no surprise that countries and companies have spent billions of dollars to modernize shipping vessels and ports with various technologies. However, the implementation of these technologies has also caught the attention of cybercriminals. For example, a cyberattack on one shipping company resulted in nearly $300 millions in financial losses. Hence, this paper describes cybersecurity vulnerabilities present in the international shipping business. The contribution of this paper is the identification and dissection of cyber vulnerabilities specific to the shipping industry, along with how and why these potential vulnerabilities exist.*

## 1. Introduction

Maritime shipping has become a trillion dollar a year industry that now impacts the economy of virtually every country worldwide (Hellenic Shipping News 2018). A problem or system failure impacting a single shipping company (for example, the 2021 Suez Canal obstruction incident) can affect businesses and organizations in a variety of industries, many of which are not aware of the criticality of the maritime shipping businesses. Further complicating matters, many countries have spent billions of dollars building and developing a large number of ports but tend to concentrate a majority of shipping trade through only a handful of ports (Tomer & Kane 2015).

Hence, it is no surprise that the shipping industry and its associated infrastructure, including shipping ports, have become high-value targets for cybercriminals and state-sponsored cyberattacks (Lagouvardou 2018; Maritime Industry National Maritime Interagency Advisory Group 2016). For example, a cyberattack on Maersk, the world's largest shipping company, resulted in nearly $300 million in financial losses (Greenburg 2018). Similarly, a malware attack

on the Mediterranean Shipping Company in April 2020 resulted in its data centre becoming unavailable for several days (Cimpanu 2020). In fact, the frequency and potential impact of cyberattacks on the shipping industry has resulted in both the European Union and the United States Department of Homeland Security (Committee on Homeland Security 2017; European Union Agency for Cybersecurity 2020) developing and publishing guidelines concerning cybersecurity countermeasures, in an effort to prevent these attacks.

The growing importance of the shipping industry, coupled with an increased number of cyberattacks results in the following research question: what are the cyber security vulnerabilities currently present within the shipping industry? While previous research has focused on specific security issues in shipping vessels, minimal research has attempted to consolidate how issues could impact the wider shipping industry, when adversaries exploit one or more vulnerability. The purpose of this paper is to answer this question through a review of relevant literature and industry reports detailing cyberattacks on the maritime shipping business. Along with academic conference and journal publications, industry reports and technical bulletins from various governmental agencies (such as the National Maritime Intelligence-Integration Office [NMIO] and the International Maritime Organization [IMO]) were examined and included in the paper. The findings from the literature review have resulted in the identification of a number of cyber security vulnerabilities in ships, ports, and associated maritime infrastructure, which are reported upon here; the primary contribution of this paper is the identification and dissection of several cyber vulnerabilities known to exist within the shipping industry, along with how and why these potential vulnerabilities exist.

The rest of the paper is structured as follows: The next section describes and discusses potential vulnerabilities identified in the literature that impact actual shipping vessels, including a ship's identification system, the Radio Detection and Ranging (RADAR) system, and bridge-based systems. The third section presents potential vulnerabilities that impact shipping ports and their associated infrastructure, including automated cranes and power systems. The fourth section presents open cybersecurity challenges for the maritime industry based on the research findings, and the final section concludes the paper.

## 2. Technologies and Cyber Threats on Vessels

As technology is increasingly being integrated into shipping vessels, many of the internal systems and computer networks on these vessels are interacting with external communication networks. As a result, there is an increased risk of unauthorized access and cyberattacks of vessels. This section describes various vulnerabilities that can be exploited involving onboard ship systems and networks, including electronic charts, RADAR, and other on-board equipment.

### 2.1 Electronic Chart Display and Information System (ECDIS)

The Electronic Chart Display and Information System (ECDIS) is the primary tool used for ship navigation and its use is mandated by the International Maritime Organization (IMO). The ECDIS relies on Electronic Navigational Charts (ENCs) from external sources, such as the National Oceanic and Atmospheric Administration (NOAA). Chart information can be loaded into the ECDIS using several different mechanisms, including external media (USB drives, DVDs, or CD-ROMs) or can be downloaded directly from the Internet.

The International Hydrographic Bureau (IHO) defines the format and structure of ENC files, defining them as standard number S-57 (International Hydrographic Organisation 2000). ENC files uploaded to a ship's ECDIS are expected to be obtained from an official IHO database (Weintrit 2018). However, ENC files are also available from third-party commercial providers, who create alternative ENCs from IHO-approved data points. These resellers are expected to independently verify the format of ENC files, so that they conform to the S-57 Standard. However, this verification process does not include validating the actual data content of the ENC file (Palikaris & Mavraeidopoulos 2020). ECDIS systems appear to accept ENC data files if they are in the correct format, regardless of the data content. As a result, a malicious insider or an individual who can obtain access to the ECDIS could, in theory, upload unauthorized ENC files to the system and could thus load malicious code that can compromise the ECDIS.

While the IHO has released a new encrypted format for ENC files (S-63 Standard), backward compatibility challenges with older vessels means that ENC files are still being released in the older S-57 format (International Hydrographic Organisation 2012). Thus, it is still possible for an attacker to create an unauthorised ENC file and to upload the file to the ECDIS. Further complicating matters, many ECDIS systems are supported by the Windows XP operating system (Xiaoxia & Chaohua 2002; Weintrit 2009). However, this operating system is known to have significant security vulnerabilities. Hence, there is the potential to replace entire data files with false information that could result in navigational errors (Bothur, Zheng & Valli 2017; Dyryavyy 2014).

## 2.2 Radio Detection and Ranging and Automatic Radar Plotting Aid

The purpose of Radio Detection and Ranging (RADAR) is to 'sweep' an area around a shipping vessel by using a steerable antenna, in order to detect objects by their reflected signature using emitted radio. While many would expect a RADAR system to involve a crew member hunched over a green circular scope, many RADAR systems also consist of a personal computer (PC) that is used to display the resulting scans from the RADAR system itself. For example, the Merlin RADAR system includes a PC that runs a version of Windows 7 Embedded edition (Kostenko 2013). As a result, RADAR systems that consist of PCs such as the Merlin option could be exploited through any of the (un)known vulnerabilities or flaws that exist within the particular operating system. Further complicating matters, unless the PC is properly locked down (Grispos, Glisson & Storer 2013), unsecured USB ports could provide an adversary with an avenue for launching an attack.

PCs running within the RADAR system could also be integrated with other sensory system units on the vessel, such as the Automatic Radar Plotting Aid (ARPA). Together with the RADAR system, the ARPA allows the ship's crew to detect and to monitor objects (for example, other ships) in real time on a map-like display (Raytheon n.d.). In addition, an ARPA system can be used by the crew to 'predict' future positions of other vessels, and therefore acts as a collision detection and avoidance mechanism. However, Marshall (1995) argues that crew members operating an ARPA system could potentially be a weakness in the overall security of the systems. Marshall states that "while almost all vessels are properly equipped as a result of vessel inspection laws, there is no similar requirement that each mariner prove his or her competence before joining a vessel" (Marshall 1995). Hence, crew members with limited or no training and, potentially, limited knowledge about the ARPA system could result in incidents that could be avoided with proper crew education.

In terms of cybersecurity vulnerabilities, Lagouvardou (2018) describes various attacks that are possible involving the ARPA system, including Distributed Denial-of-Service (DDoS) attacks, intentional jamming of the signal, and the spoofing or modification of signals reflected back to the system from objects that are not really in the position detected by ARPA. Furthermore, if a ship does not integrate Global Positioning System (GPS) into the ARPA system, it may be vulnerable to "display[ing] target information oriented along the ship's heading, the so-called heads-up mode, with relative motion" (Bhatti & Humphreys 2017). This problem is further amplified by the number of vulnerabilities that exist within GPS systems (Adams 2001; Achanta *et al.* 2015).

## 2.3 Automatic identification systems of vessels

The Automatic Identification System (AIS) allows maritime authorities and other ships to track and monitor vessel movements in the same way that an airport's control tower monitors aircraft (Bhattacharjee 2019). Within AIS, each vessel is assigned a unique identifier that other ships can use to track their location. This information can also be used to locate ships when search and rescue missions are needed in the event of an incident or accident onboard the ship. As a result, the International Maritime Organisation (IMO) mandates that AIS is fitted to all passenger ships that travel internationally as well as to any ships that travel through international waters and weigh over 300 gross tonnes (Meng, Weng & Li 2014).

However, AIS is not without its potential security issues. Many ships frequently implement AIS on the same computer as an Electronic Chart Display and Information System (ECDIS). As a result, the same hardware and software challenges that exist within ECDIS also apply to AIS. Balduzzi, Pasta & Wilhoit (2014) identified several potential vulnerabilities in AIS, including a lack of authentication, the ability to spoof Aids to Navigations (AtN) messages, the triggering of collision alarms, luring vessels via fake search-and-rescue signals, man-in-the-middle attacks, as well as denial-of-service attacks. Another potential vulnerability exists with the approach used by AIS providers to forward and transmit AIS data. This forwarding process uses little or no authentication or encryption and transmits information over Internet Protocol (IP) UDP Port 5321. These AIS providers do not validate incoming sources of the AIS data; as a result, an adversary could inject or spoof malicious AIS data if the hosting network can be accessed.

## 2.4 Maneuvering systems

The overwhelming majority of shipping accidents can be attributed to human error (Celik & Cebi 2009; Weng et al. 2019). As a result, many ships are choosing to deploy safety-enhancing technology in order to reduce errors and improve overall efficiency. One piece of technology that has greatly assisted ship crews is a maneuvering system. Undertaking a manoeuvre in high-traffic areas, such as in ports or in coastal waters, is inherently more dangerous, which increases the likelihood of an accident occurring. For example, there have been several high-profile cases of ships striking the Oakland Bay Bridge near San Francisco in the United States (Rogers 2013; Taylor, Nolte & Curiel 2007).

Hence, the introduction of new technology has assisted with these types of manoeuvres. For example, the Deep-Sea Navigation System (DSNS) is an autopilot system that assists a ship crew to ensure they follow their planned route, but still allows for changes based on severe weather or complex traffic situations. This system also allows for remote onshore pilots to control the ship, if the need arises (Viljanen 2020). While there are minimal details on vulnerabilities affecting DSNS, this system utilises GPS signals in order to provide accurate ship position information. Therefore, any vulnerabilities that impact GPS systems will also impact DSNS, and the maneuvering system could, potentially, be fooled into an erroneous path.

## 2.5 Global positioning systems

Many systems within a shipping vessel rely directly or indirectly on the Global Positioning System (GPS) for accurate location and timing information (Nguyen 2020; Park & Kim 2011). GPS consists of three main components: a constellation of 24 satellites, a control segment that maintains the satellite orbits and general upkeep, and the GPS receivers that are used to determine location and time (Larcom & Liu 2013). In order for a GPS unit to determine its location, it must have a line-of-sight signal to at least three satellites. However, GPS receivers do not directly 'look up' for a satellite, but instead focus in an omnidirectional manner. As a result, GPS receivers are susceptible to spoofing attacks.

For example, Tippenhauer *et al.* (2011) demonstrated how a group of GPS receivers can be fooled by transmitting a strong enough signal that appears to come from three satellites. Larcom and Liu (2013) describe a more complex attack called a 'satellite lock takeover' that is similar to man-in-the-middle attack. In this scenario, legitimate GPS signals are received and then rebroadcast with modified coordinates. As a result, the GPS receiver gradually 'drifts' off and ends up in a position that is offset from its actual location.

Unfortunately, due to the availability of detailed online documentation and equipment that can be purchased (albeit illegally) on eBay, GPS spoofing is well within the technology capabilities of amateur malicious actors (Hill 2017; Huang & Yang 2015). While several mitigation techniques have been described in the literature (Ioannides, Pany & Gibbons 2016; Larcom & Liu 2013), it appears that the adoption of these techniques is not widespread in the shipping industry.

## 2.6 Notice to mariners corrections

On a shipping vessel, an extremely important necessity is for the crew to have access to up-to-date correction charts showing the physical location of obstructions, sea depths at certain locations, dangerous areas to travel, and course-plotting assistance markers. These data points change from time to time and need to be updated periodically so as to safeguard the vessel while in passage. In the United States, information related to coastlines and harbours that are within the U.S. or its territories are published by the National Ocean Service. Other nations provide changes to the National Geospatial-Intelligence Agency for waters outside the U.S. (Lee 2007).

In terms of cyber threats to this technology, correction chart updates are published on a typical website, accessible by the general public (The National Oceanic and Atmospheric Administration 2019). As a result, these chart updates are susceptible to a variety of web attacks, which include redirecting ship crews to fraudulent websites that contain malicious binary code embedded within fake correction chart updates. Without a means to determine if an update is authentic, a ship's crew could update its charts with incorrect information, potentially resulting in the vessel striking other objects either in the ocean or on land.

## 2.7 Global Maritime Distress and Safety System (NAVTEX)

In 1979, the International Maritime Organization (IMO) initiated the Global Maritime Distress and Safety System (GMDSS) as a way of establishing a set of policies and procedures intended to increase the safety of shipping vessels (Kent 1989). The procedures encompass specific equipment and communication protocols, which together enable the timely search and rescue of vessels. The policies apply to vessels that weigh over 300 tonnes and require ships to contain a communication system that consists of either a radio or satellite, which can be used to send or to receive distress calls.

Another feature of the GMDSS is the integration of the Navigational Telex (NAVTEX) system, which receives automated maritime safety information. NAVTEX consists of a set of safety procedures and protocols, as well as equipment that is used to provide maritime safety information to other ships. This information includes detailed weather information and any relevant warnings (Bauk 2019). This allows groups and agencies responsible for search and rescue missions to have information at their disposal in order to coordinate the rescue operation.

NAVTEX messages are transmitted using the open standard Simplex Teletype over Radio (SITOR-B) (Federal Register 1995). While radio teletype has been around since the 1950s and has been extended to include error correction in the 1970s, there is no inherent security with these message types. This means that it is possible to construct a properly formatted message and broadcast it to all ships in a specific area. These messages could contain false weather information, or a notice to avoid particular areas, causing a ship to change course. One potential mitigation strategy could be the corroboration of this information through other means, such as web portals.

## 2.8 Dynamic Vessel Positioning

In certain scenarios, shipping vessels 'hover' over a particular location in the ocean. This is usually accomplished using a system called Dynamic Positioning (DP) that maintains the vessel location relative to some stationary object (Sørensen 2011). While the DP system is frequently used for offshore drilling vessels, passenger ships use the same system to prevent 'corkscrewing' in the water. The IMO defines various safety levels according to the needs of the dynamic positioning system. These levels range from level one (no redundancy) to level three (extensive redundancy) (Kaushik 2016). To maintain position, a ship can use a combination of approaches including 'Light Taught Wire' – a simple weight suspended below the ship, different GPS systems, a gyrocompass, as well as seabed acoustics. These various systems provide information to a computer, which then produces a measured model of the ship, taking into consideration the current, wind, drag of the vessel, the ship's location, and power of the thrusters being used.

One vulnerability associated with the DP system is the spoofing or modification of the GPS signal used for the calculations (Grant *et al.* 2009). Bhatti and Humphreys (2017) add that, because the DP system relies heavily on GPS, an attack that focuses on gradually drifting the GPS signal, the vessel's position will shift in a way which may not be detected, but instead be blamed on the wind or a change on the ocean current. It would also appear that certain DP systems have been found to contain malware; one incident reported in the literature describes an oil drilling rig where the dynamic positioning system contained a virus which was  undetected for nearly nineteen days (Bhatti & Humphreys 2017). Lastly, a regulator in the United Kingdom reported that it was possible for malicious actors to perform a denial of service attack on DP systems that will impact a ship's dynamic positioning (United Kingdom Health and Safety Executive 2013).

## 2.9 Other Vessel Communication Systems

In addition to the communications methods discussed above, shipping vessels also use a variety of other communication technologies. For example, the radio systems on many ships make use of UHF/VHF, S-Band, and even cell technologies—such as GSM, 3G and 4G (Tian *et al.* 2017). Another example is the implementation of Private Automated Branch Exchange (PABX) technologies in order to allow on-board data communications, including voice over IP, within ships. Many of these onboard networks consist of off-the-shelf components that increase the potential for malicious insiders (for example, crew members) to introduce malware via removable media.

Satellites also appear to be a popular transmission medium within shipping vessels (Rosa, Meidenbauer & Halla 2016). The most commonly used satellite systems used include Irridium, Inmarsat, VSat, Broadband Global Area Network, and FleetBroadband (Santamarta 2014). However, satellite-based networks are susceptible to uplink and downlink jamming, where a malicious actor floods a signal, transmitter, or receiver with 'noise' with the purpose of interfering with the legitimate transmission (Hudaib 2016).

## 3. Technologies and Cyber Threats at the Shipping Port

In addition to vulnerabilities existing in many modern shipping vessels, cyber threats are also becoming increasingly visible in shipping ports around the world. This section introduces cyber threats that involve port systems, including crane systems, power systems, as well as physical security systems. The first vulnerability examined is crane systems, and these as well as other port-side equipment have vulnerabilities similar to those onboard vessels.

### 3.1 Crane systems for offloading and onloading

Standard containers in the cargo industry consist of a steel or aluminium metal box with explicit dimensions and loading/unloading anchor points (InterPort 2020). The majority of cargo is transported in these standard units, but specialized containers might also include liquid containers or refrigerated units. Because of their standardized dimensions, the loading and unloading of containers from a ship is undertaken by robotic cranes, together with the assistance of Global Positioning Systems (GPS) (Gotting 2001). Also, with increased automation, one crane operator may be operating several cranes, relying on automation to finish moving the individual containers. These heavily automated cranes are responsible for the quick loading and unloading as vessels come in or leave a port. The primary benefit of using GPS for this operation is to reduce the number of misplaced containers.

However, crane GPS systems are increasingly susceptible to GPS spoofing. The consequences of these attacks range from a robotic crane loading the incorrect container, placing a container on the wrong ship, or even worse, dropping the container into the sea or ocean. Moreover, the increased automation of robotic cranes means modern crane systems also communicate with a variety of off-the-shelf port-side systems. This introduces a host of vulnerabilities and potential points of access for malicious actors to exploit entire crane ecosystems (Limer 2019).

### 3.2 Port power systems and vulnerabilities

A port's electrical power and energy infrastructure are critical to ensuring continuous operations as well as the safety of port workers, customers, and the goods traveling through the port. However, given that every post is different in terms of size, scope, and the nature of the traffic that flows through it, it is difficult to generalize the cybersecurity risks that a specific port might face. This is especially true when discussing today's electrical and power infrastructure that could contain both physical and remote components, known as cyber-physical systems *(Holmgren*

Jenelius & Westin 2007; ICF International 2016). As a result, it is critical to make sure the physical infrastructure and the underlying electrical systems are protected and secure. Systems that need to be secure include instrument transformers, insulators, isolators, smart feeder switches, smart meters, and micro-grid controllers (Davis 2009). Other systems that are also susceptible to cyberattacks include Energy Storage Systems, which include large battery banks to help power port equipment as well as Alternate Maritime Power that provides shore-to-ship power delivery.

Power and electrical systems once again rely on GPS systems. For example, a device known as a synchrophaser uses GPS timing to synchronize alternating current in an electrical power grid (Shepard, Humphreys & Fansler 2012). For instance, some electrical power might be supplied by hydroelectric plants, while other electricity is supplied by wind or solar options. Since the alternating current must be coordinated from all of the power sources on the grid, the synchrophaser uses GPS timing to adjust the alternating current to the correct phase. Thus, GPS spoofing can potentially impact the supply of power to the port. This vulnerability has already been reported (Staff 2012), but, at the time of writing, no such attack has been identified in the real world.

Moreover, vulnerabilities also exist within the power system's information communication and industrial control systems (Martin & Reynolds 2016). These are systems used to enable disparate systems to exchange information and to control the underlying physical hardware. According to the National Renewable Energy Lab in the United States, these devices are a concern with regard to cyberattacks (Martin & Reynolds 2016). The main cause of concern for these devices stems from the embedded operating system they run on as well as the way in which they communicate with other parts of the system.

## 3.3 Physical and access security

While not always directly applicable to cybersecurity incidents or digital crimes, the physical security of the port nonetheless intersects with the world of cyberattacks. The United States Department of Homeland Security (DHS) Security Grant Program was created in part to address physical port security, including the acquisition of control measures: gates, guns for guards, and ID cards with access codes (Federal Emergency Management Agency n.d.). In recent years, the grant has been extended to include cyber threats, with DHS making the enhancement of cybersecurity capabilities a priority within the program (United States Department of Homeland Security 2020). The hope is to encourage port facilities to take the steps necessary to increase the security of their critical information communication systems. Potential vulnerable points of attack from a physical aspect include connections to computer networks, the Internet, remote communications, control and monitoring, and Supervisory Control and Data Acquisition (SCADA) devices. All of these devices have been discussed in the literature as containing multiple vulnerabilities (Igure, Laughter & Williams 2006).

## 4. Open Challenges

As discussed in previous sections, current issues in the area shipping vessel and port security support the development of research agendas to address these issues. This section specifically examines three areas of research that help improve and enhance the cybersecurity of the shipping industry, including the promotion of security-by-design shipping vessels, the development and advocacy of specific cybersecurity standards for the shipping industry, and an understanding of incident response/cyber investigations of shipping vessels.

## 4.1 Security-by-design shipping vessels

The idea behind security-by-design is that developers and designers take security into account from the early stages of the design or development process (Casola *et al.* 2016). One potential solution to the growing number of cyber threats to the maritime industry could thus involve the design and development of ships and bridge equipment which prioritize cybersecurity requirements. This would involve integrating requirements for security into relevant phases of the ship's design and development, with the objective of producing a ship that can be considered secure-by-design.

As evident by the above discussion, a typical ship's architecture could consist of different layers and could contain multiple types of devices. For example, consider the implementation of Radio Detection and Ranging (RADAR) systems that consists of hardware that includes USB ports, which can be exploited to launch malware attacks. In a security-by-design approach, designers and ship developers would examine a RADAR system's attack surface, identifying how the system can be exploited, and then identify security controls to mitigate or remove the risks posed by the identified threats (Geismann, Gerking & Bodden 2018). In this simple example, one potential solution could involve 'locking' down the USB ports, effectively removing the option for an attacker to use the ports to launch executable malware. Another potential vulnerability that could be mitigated with a security-by-design approach is the approach used by AIS providers to forward and transmit AIS data, which uses little or no authentication and encryption and transmits information. Again, system designers could identify that this data transmission mechanism is open to abuse by an adversary who can inject or spoof AIS data. By identifying that this security flaw exists (as part of the attack surface), designers and developers can mitigate the risk by implementing a message validation mechanism.

## 4.2 Specific cybersecurity standards for the shipping industry

While the above discussion has identified several cybersecurity challenges in the shipping industry, it appears that the industry has been slow to adopt and implement standards and best practices with regard to cybersecurity. While the situation is improving, there is still room for improvement. For example, the International Maritime Organization (IMO) has taken the lead on publishing cybersecurity regulation concerning the shipping industry called "Guidelines on Maritime Cyber Risk Management" (International Maritime Organization n.d.). In addition, the IMO has also published a resolution entitled "Maritime Cyber Risk Management in Safety Management Systems" as well as penalties for (or lack of) compliance (International Maritime Organization n.d.). However, the guidelines only provide "high-level recommendations", while the resolution, simply "encourages administrations to ensure that cyber risks are appropriately addressed" and only took effect in January 2021. One potential obstacle towards the implementation and enforcement of shipping cybersecurity standards is that the industry spans countries, companies, requirements, and ports. As a result, the IMO would need to have the leverage to enforce the actual rules. Hence, short of an actual authority with the ability to penalize for compliance, the situation in terms of security standards may not be changing any time soon.

## 4.3 Incident response and cyber investigations for shipping vessels

Traditionally, when organisations detect or identify a security incident or digital crime, one response is to conduct an incident response/cyber investigation (Grispos, Glisson & Storer 2019,

Grispos *et al.* 2017). The purpose of this investigation is to collect and analyse data from systems affected by the incident/crime, in order to order establish the cause and how it can be prevented in the future. In terms of the shipping industry, the IMO's "Guidelines on Maritime Cyber Risk Management" provide high-level recommendations that maritime stakeholders can take in the event a security incident or digital crime occurs. These guidelines include implementation of activities to detect the cyber event, the development of activities and plans to restore systems for normal shipping operations, as well as measures to recover and resume shipping operations as quickly as possible after a cyber event (International Maritime Organization n.d.). The IMO guidelines go on to suggest that maritime stakeholder consult "relevant international and industry standards and best practices such as ISO/IEC 27001 and National Institute of Standards and Technology's (NIST) Framework for Improving Critical Infrastructure Cybersecurity" for further information (International Maritime Organization n.d.).

The problem is that both the ISO/IEC 27001 and NIST standards have been designed and developed with traditional computer systems in mind, and do not take into consideration the complexity of maritime vessels. For example, NIST's 800-61 Computer Security Incident Handling Guide (National Institute of Standards and Technology 2012) suggests that when incidents or crimes occur, impacted systems should be isolated, potentially powered down, and data collected in order to conduct an investigation; in a shipping vessel situation, this might be difficult or even impossible to achieve. Further complicating matters, there is the potential that tools and techniques that are recommended for use by the incident response/cyber investigation community are invalidated in shipping vessel contexts. For example, incident handlers are guided to use tools such as 'dd' and 'FTK Imager' to create complete bit-for-bit copies of storage devices under investigation (Grispos*,* Storer & Glisson 2012). These tools are unlikely to work with 'storage devices' in shipping vessels and obtaining a complete bit-for-bit copy could be both dangerous from a safety perspective and cumbersome. Hence, the maritime industry requires guidelines, tools, and techniques specific for situations that involve shipping vessels and their respective system architectures.

## 5. Conclusions

This paper attempted to answer the research question: *what are the cyber security vulnerabilities currently present within the shipping industry*? The integration of technology into shipping vessels and ports is creating a scenario that is encouraging security incidents and digital crimes. The number of systems involved, their complexity but accessibility, the open standards involved, and many additional factors all contribute to the acceleration of potential cyber incidents. These vulnerabilities exist due to the use of antiquated standards and equipment, the lack of a security mentality when shipping systems are designed, and the availability of capable exploits which take advantage of these shortcomings. This paper has highlighted several critical areas of concern for the shipping industry that need to be addressed. However, the response to these cybersecurity concerns must start from the top, effectively beginning with organisations such as the International Maritime Organisation (IMO) which should look to develop more detailed and enhanced cybersecurity guidelines specifically for the shipping industry. But guidelines by themselves are insufficient. The importance of the shipping industry on global economies should force legal requirements that should be compulsory on shipping manufacturers, shipping companies, crew members, and port providers. Hopefully, this paper will start the conversation and will provide the basis for future discussions regarding the need for enhanced cybersecurity measures in the shipping industry.